# An Efficient Adaptive Boundary Matching Algorithm for Video Error Concealment

S. M. Marvasti-Zadeh*, H. Ghanei-Yakhdan* and S. Kasaei**[(C.A.)]

**Abstract:** Sending compressed video data in error-prone environments (like the Internet and wireless networks) might cause data degradation. Error concealment techniques try to conceal the received data in the decoder side. In this paper, an adaptive boundary matching algorithm is presented for recovering the damaged Motion Vectors (MVs). This algorithm uses an outer boundary matching or directional temporal boundary matching method to compare every boundary of candidate Macro-Blocks (MBs), adaptively. It gives a specific weight according to the accuracy of each boundary of the damaged MB. Moreover, if each of the adjacent MBs is already concealed, different weights are given to the boundaries. Finally, the MV with minimum adaptive boundary distortion is selected as the MV of the damaged MB. Experimental results show that the proposed algorithm can improve both objective and subjective quality of reconstructed frames without any considerable computational complexity. The average PSNR in some frames of test sequences increases about 5.20, 5.78, 5.88, 4.37, 4.41, and 3.50 dB compared to average MV, classic boundary matching, directional boundary matching, directional temporal boundary matching, outer boundary matching, and dynamical temporal error concealment algorithm, respectively.

## 1 Introduction

Because of enormous amount of digital video data and limited bandwidth of communication channels, the original raw data needs to be compressed. Data compression is mainly obtained by removing the spatial, temporal, and psychovisual redundancy; which is disregarded by Human Visual System (HVS). However, removing data redundancies makes it to become highly sensitive to channel issues; such as noise and congestion. The lost video data through communication channel decreases the quality of received data.

Channel errors in received video frames might propagate through time (because of motion compensation techniques). On the other hand, error resilient (such as Unequal Error Protection (UEP) [1]) and Automatic Retransmission Request (ARQ) methods increase the requested bandwidth and delay the video data transmission. Also, these methods decrease the coding efficiency.

Error concealment is a useful mechanism which conceals the received damaged data in the decoder side without increasing the redundancy of sending sequences and the required bandwidth. These techniques are divided into two main types: spatial domain [2-5] and temporal domain techniques.

The spatial domain error concealment techniques try to recover the damaged data by using the correctly received data in adjacent damaged areas in the same frame. The temporal domain error concealment techniques recover the damaged area in video frames by using the temporal correlation between consecutive frames. In this paper, we have focused on temporal error concealment.

The simplest temporal error concealment method is the Temporal Replacement (TR) [6]. In that method, all damaged MVs are replaced by zeros. The method is useful when there is a low motion among consecutive frames. In [7], a whole frame loss error concealment algorithm is proposed to further refine the TR. There are also some other simple methods which have special usages. One of these methods [8] is the use of corresponding MB's MV from previous frame, which has the better performance. When the motion in video sequences is smooth, this method is useful.

Another simple and common method is the use of the average or median of adjacent MVs of damaged MB [9]. The simulation results show that the use of median is better than average method. In [10], the performance of five conventional temporal error concealment methods in critical situations are analyzed with two similarity metrics.

In MV interpolation method [11], the MV of each block 4×4 is estimated by interpolation from MVs of adjacent MBs. The distance between adjacent blocks is used as the weight.

The Lagrange Interpolation (LI) method [12] is also a simple and useful method for MB recovering of 4×4 blocks. It supposes that the movement tendency of damaged MV is similar with its adjacent MVs. The results of that method will not be satisfactory if this supposition is not true.

One of the most conventional algorithms for temporal error concealment is the classic Boundary Matching Algorithm (BMA) [13]. It supposes that the luminance differences in boundaries of damaged MB change smoothly. Therefore, if the correct motion is selected the changes from boundary pixels of adjacent MBs are negligible. However, this algorithm has noticeable limitations. For instance, when all of the adjacent boundaries do not exist or the adjacent pixels have sudden changes, the performance will be decreased.

In [14], a fast and adaptive boundary matching algorithm is presented. At first, the method selects an adaptive search area based on the maximum repeated MV around the lost MBs. Then, it chooses an adaptive template width (the thickness of pixels around the lost MBs) depending on the smoothness of adjacent MBs.

In [15], a Kalman-filter-based error concealment technique for BMA is proposed. The idea is to model the problem of error concealment with an appropriate state-space variable and Kalman filter. In that technique, the estimated MVs by BMA are corrected by Kalman filter.

The Edge Adaptive Boundary Matching Algorithm (EA-BMA) [16] uses a mask adaptive to the edge strength in adjacent regions and boundary regions of the damaged block. It supposes that features of the damaged region and its boundary are continuous. The BMA-based algorithm employs the masks outside a damaged block instead of predicting the boundary pixels of the damaged block for estimating of the best MV.

In [17], a novel boundary matching algorithm for temporal error concealment of video frames has been proposed. It uses two boundary matching criterions (namely classic Boundary Matching Criterion (BMC) and Proposed Boundary Matching Criterion (PBMC)) to identify matching distortion at each boundary of the candidate MB, adaptively. This algorithm uses additional data for improving the accuracy of MV estimation. It introduces additional boundaries, which is exploited from the reference frame. These boundaries include the outer boundaries of adjacent MBs from the damaged MB that are overlapped with inner boundaries of damaged MB. In addition, it determines the priority list of error concealment for each video frame. The list is updated after reconstruction of each damaged MB. The list leads to increase the accuracy of MVs estimation in next stages of error concealment process.

The effective temporal error concealment algorithm [18] constructs a limited candidate MV set among the MVs of neighboring MBs and extrapolates MVs, adaptively. It uses temporal-spatial correlation of MVs to reduce the number of candidate MVs, effectively. Finally, it selects the best MV from the limited candidate MV set by using the BMA to conceal the corrupted MB.

The error concealment technique with block boundary smoothing [19] improves the subjective video quality by compensating the discontinuities at lost MB boundaries. It uses the Flexible MB Ordering (FMO) technique. In addition, it uses the weighted boundary pixels of the reference block to decrease the blocking artifacts problem.

Huang and Lien [20] proposed an adaptive temporal error concealment technique using a Self-Organizing Map (SOM). To overcome the disadvantage of the BMA, it uses a SOM as a predictor to estimate the MVs of damaged MBs more accurately. The estimated MVs are utilized to reconstruct the damaged MB by exploiting the spatial information from reference frames; based on the boundary matching criterion.

The dynamic temporal error concealment technique using a Competitive Neural Network (CNN) [21] employs a CNN predictor or the BMA for estimating the MV of damaged MB. Based on video scene motion, the technique uses different methods. Therefore, it has low computation complexity and is suitable for real-time applications.

The fuzzy reasoning-based temporal error concealment method [22] uses two measuring criterions (namely Side Match Distortion (SMD) and Sum of Absolute Difference (SAD)) together for reflecting the matching status of candidate MBs. Subsequently, the fuzzy reasoning is adopted to balance the effects of two criterions to accomplish the judgment more accurately for candidate MVs. Also, *Lai et al.* [23] proposed a Fuzzy Metric Based Boundary Matching Algorithm (FBMA) which uses Sugeno fuzzy integral as the criterion to compare the candidate MVs. The reconstructed frames with this criterion are more consistent with the HVS.

In [24], a method for MV optimization of damaged MB is presented with the two best MVs that are obtained from BMA. Furthermore, a pre-processing step is presented for determining a better MVs set.

According to BMA criterion with reliability coefficient, in double weighted MVs algorithm [25] the two best MV are selected. Then, the optimal MV is calculated by weighting the MVs in terms of their accuracy.

The efficient MVs interpolation method [26] interpolates the MVs of lost blocks in the current frame with the help of extrapolated MVs from the previous frame. It increases the accuracy in recovering the corrupted block MVs compared with conventional methods.

The enhanced edge-sensitive processing order algorithm [27] uses a suitable processing order and a new MV searching algorithm for temporal error concealment. It uses an efficient processing order for error concealment by considering the edge and MV information of neighboring MBs. Finally, the best candidate MV is chosen by the modified Texture-Based Selective Block Matching Algorithm (TSBMA).

An adaptive error concealment mechanism based on decision tree [28] uses temporal or spatial methods depending on instantaneous features of video sequence frames and the error pattern.

An integrated temporal error concealment technique for H.264/AVC [29] utilizes adaptively integrating error concealment approaches with the adaptive weight-based switching algorithm. It adaptively switches between two modes. The first mode is a conventional temporal error concealment mode and the second mode is an integrated mode that is based on spatial evaluation criteria. In addition, the second mode is obtained by integrating two temporal error concealment approaches with an adaptive weight. It can obtain the optimal recovery data for damaged MBs.

Wu et al. [30] proposed an efficient spatial-temporal error concealment algorithm for H.264/AVC with a low computational complexity. It uses a frame-level scene-change detection to decide whether spatial or temporal error concealment should be applied. For temporal error concealment, a low complexity prediction-based MV estimation scheme is applied for estimating the best MV.

The Outer Boundary Matching Algorithm (OBMA) [31] uses a linear translational model for error concealment of damaged MB. In addition, in [32] a low-complexity error concealment technique with the use of OBMA is investigated for mobile applications. In the matching process, the criterion function of OBMA gives equally importance to boundaries of adjacent MBs. It caused low accuracy in the process, because, there will be error concealment in some of adjacent MBs from previous stages. So, if there is not enough accuracy in adjacent MBs, there will not be accurate computed MVs from OBMA. This caused error-propagation in estimating of the next damaged MBs. Also in [33], a Dynamic Temporal Error Concealment (DTEC) method for estimating damaged MVs was presented. Depending on the motion regions of each frame, the method uses one of the TR, adjacent MV, or the improved OBMA.

A Spatio-Temporal Boundary Matching Algorithm (STBMA) [34] estimates the damaged MVs with the use of BMA and OBMA criterions and side smoothness criterion of damaged MB. Also, an Efficient Spatio-Temporal Boundary Matching Algorithm (ESTBMA) [35] recovers the lost MVs by minimizing a distortion function which exploits both spatial and temporal smoothness properties.

Most conventional error concealment methods (such as BMA) apply only one direction to calculate the differences among boundaries in boundary distortion function. The Directional Boundary Matching (DBM) method [36] uses a simple edge detection process to improve the accuracy of damaged MV estimation. First, it determines the direction of comparison for each boundary pixel of the candidate MB. Then, every boundary pixel in the edge direction is compared with a pixel of outer boundary of the damaged MB.

The Directional Temporal Boundary Matching Algorithm (DTBMA) [37] estimates the real boundary direction by considering pixel differences in three directions for temporal error concealment. If candidate MBs are from previous stages of error concealment, the DBM and DTBMA will not lead to satisfactory results. In general, the DTBMA has a better performance compare to BMA and DBM.

To solve the problems of conventional algorithms, the proposed algorithm with adaptive calculation of boundary distortion function, estimates the damaged MV. In this paper, an adaptive boundary matching algorithm for error concealment of video frames is presented. The distortion of outer boundary matching and the distortion of directional temporal boundary matching are calculated and then just one of them is selected for each boundary, adaptively. Finally, the best MV is estimated by minimizing the distortion function for all boundaries of the damaged MB.

Moreover, according to accuracy of each outer boundary from damaged MB, there is a specific weight for each boundary. It also gives different weights to each adjacent boundaries of damaged MB which is the result of previous error concealment stages, adaptively. These weights increase the accuracy of estimated MVs and decrease the error propagation effects.

The rest of this paper is organized as follows. Section 2 describes the proposed algorithm in detail. The experimental results are given in Section 3. Finally, conclusions are summarized in Section 4.

## 2 Proposed Error Concealment Algorithm

The proposed error concealment algorithm uses two boundary matching criterions for selection of candidate MBs' boundaries with adjacent MBs' boundaries. The first criterion is the outer boundary matching method which compares the boundary pixels without determining the direction of edge on each boundary. The second criterion is the directional temporal boundary matching method, which determines the selected direction of each boundary pixels. It then compares the boundary pixels in determined directions. Of course, in these two criterions, it gives a specific weight to each boundary in order to increase the accuracy and prevent error propagation. In this work, in order to focus on the problem of error concealment, it is assumed that the positions of damaged MBs in video frames are known. For error detection techniques, the interested readers can refer to [38-40].

### 2.1 Outer Boundary Matching Criterion (OBMC)

This criterion is mainly the same as the used distortion function criterion in OBMA. The difference between them is that according to the accuracy of each adjacent boundary from damaged MB, there is a specific weight to this boundary. This criterion makes use of spatial and temporal smoothness on boundaries of damaged MB.

If consecutive frames have a low temporal correlation, the efficiency of the criterion decreases. OBMC function for each boundary is calculated by Eq. (1) to Eq. (4), respectively.

$$OBMC_{top} = w_{top} \times \sum_{n=0}^{S-1} | f_{cur}(i+n, j-1) - f_{ref}(i+vi+n, j+vj-1) | \quad (1)$$

$$OBMC_{bottom} = w_{bottom} \times \sum_{n=0}^{S-1} | f_{cur}(i+n, j+S) - f_{ref}(i+vi+n, j+vj+S) | \quad (2)$$

$$OBMC_{left} = w_{left} \times \sum_{n=0}^{S-1} | f_{cur}(i-1, j+n) - f_{ref}(i+vi-1, j+vj+n) | \quad (3)$$

$$OBMC_{right} = w_{right} \times \sum_{n=0}^{S-1} | f_{cur}(i+S, j+n) - f_{ref}(i+vi+S, j+vj+n) | \quad (4)$$

where $(i,j)$ is the location of the top-left pixel in the damaged MB, $f_{cur}(.,.)$ stands for current frame, $f_{ref}(.,.)$ is the corresponding reference frame, $MV(vi,vj)$ shows the candidate MV, and $S$ indicates the number of pixels on each boundary. Also, in Eq. (1) to Eq. (4), *top*, *bottom*, *left* and *right* are referred to top, bottom, left, and right adjacent MB, respectively. Moreover, $w$ is the weight of each outer boundary from damaged MB. How to determine weights of each outer boundary from damaged MB is explained in Subsection 2.4, separately. Outer boundaries of damaged MB and candidate MB are shown in Fig. 1.

### 2.2 Directional Temporal Boundary Matching Criterion (DTBMC)

This criterion is mainly similar to the used criterion in DTBMA. Like OBMC, this criterion uses the specific weights for adjacent boundaries from damaged MB. When the direction of edges on boundaries from damaged MB has not been changed (compared to the last frame) this criterion works well. However, in some cases, the edge direction determination according to boundary pixels of candidate MB can cause low accuracy in MV estimation. While the boundary edges' direction of damaged MB is changed with respect to the last frame, this criterion will not lead to acceptable results. Moreover, if the candidate MB is from previous stages of error concealment, the efficiency of criterion decreases. Also, DTBMC uses the spatial and temporal smoothness feature on damaged MB boundaries to find the best candidate MV.

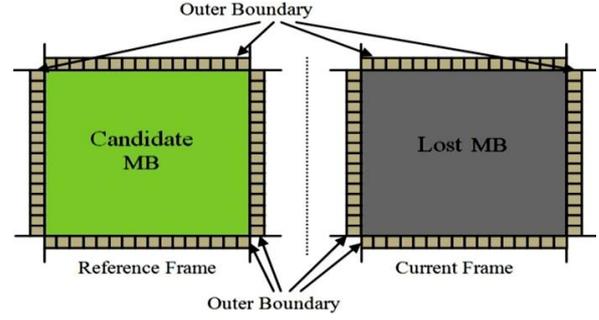

**Fig. 1** Illustration of outer boundaries in OBMC.

At first, a correct direction is identified for each boundary pixel of the candidate MB. This direction is identified by inner and outer boundaries of candidate MB. Then, each boundary pixel of the candidate MB is compared with the identified direction of outer boundary pixels of damaged MB. Thus, the estimation of damaged MV is more accurate.

Fig. 2(a) shows the edge direction identification method for each top boundary pixel in the reference frame in DTBMC. Also, Fig. 2(b) shows how to compare them with boundary pixels of the current frame in DTBMC.

The boundary matching process for estimating of damaged MB using DTBMC is as follow. At first, the correct direction for comparing each $l^{th}$ pixel in $(i,j)$ coordinate in each boundary is calculated by

$$E^l_{dir} = \min(E_{l-1}(dir), E_l(dir), E_{l+1}(dir))$$
$$dir \in \{top, left, bot, right\} \quad (5)$$

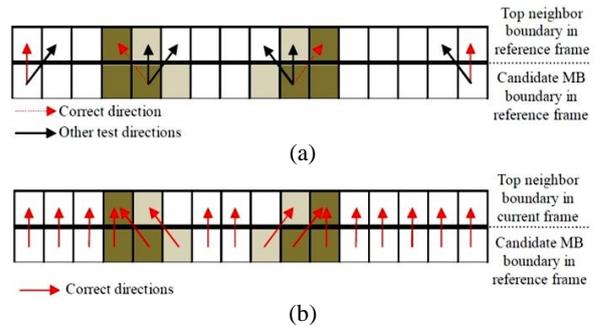

**Fig. 2** Illustration of: (a) identifying comparison direction for each boundary pixel (b) applying matching process in determinate directions.

where $E_{dir}^{l}$ is the determinant of correct direction for comparing of $l^{th}$ pixel. The $l^{th}$ pixel can be in each top, left, bottom, or right boundary. Also, $E_{l-1}(...)$, $E_{l}(...)$, $E_{l+1}(...)$ are absolute difference values between $l^{th}$ pixel and $l\text{-}1^{th}$, $l^{th}$, and $l\text{+}1^{th}$ pixels in outer boundary of the candidate MB, respectively.

These differences are calculated in top, left, bottom, and right boundary by Eq. (6) to Eq. (9), respectively.

Then, the boundary pixel of the candidate MB in the identified direction is compared with the outer boundary pixel of the damaged MB by Eq. (10) to Eq. (14), respectively,

$$\begin{bmatrix} E_{l-1}(top) = |f_{ref}(i+vi, j+vj) - f_{ref}(i+vi-1, j+vj-1)| \\ E_{l}(top) = |f_{ref}(i+vi, j+vj) - f_{ref}(i+vi, j+vj-1)| \\ E_{l+1}(top) = |f_{ref}(i+vi, j+vj) - f_{ref}(i+vi+1, j+vj-1)| \end{bmatrix} \quad (6)$$

$$\begin{bmatrix} E_{l-1}(left) = |f_{ref}(i+vi, j+vj) - f_{ref}(i+vi-1, j+vj-1)| \\ E_{l}(left) = |f_{ref}(i+vi, j+vj) - f_{ref}(i+vi-1, j+vj)| \\ E_{l+1}(left) = |f_{ref}(i+vi, j+vj) - f_{ref}(i+vi-1, j+vj+1)| \end{bmatrix} \quad (7)$$

$$\begin{bmatrix} E_{l-1}(bot) = |f_{ref}(i+vi, j+vj+S-1) - f_{ref}(i+vi-1, j+vj+S)| \\ E_{l}(bot) = |f_{ref}(i+vi, j+vj+S-1) - f_{ref}(i+vi, j+vj+S)| \\ E_{l+1}(bot) = |f_{ref}(i+vi, j+vj+S-1) - f_{ref}(i+vi+1, j+vj+S)| \end{bmatrix} \quad (8)$$

$$\begin{bmatrix} E_{l-1}(right) = |f_{ref}(i+vi+S-1, j+vj) - f_{ref}(i+vi+S, j+vj-1)| \\ E_{l}(right) = |f_{ref}(i+vi+S-1, j+vj) - f_{ref}(i+vi+S, j+vj)| \\ E_{l+1}(right) = |f_{ref}(i+vi+S-1, j+vj) - f_{ref}(i+vi+S, j+vj+1)| \end{bmatrix} \quad (9)$$

$$E_{l}^{top} = \begin{bmatrix} |f_{ref}(i+vi, j+vj) - f_{cur}(i-1, j-1)| & \text{if} \quad E_{dir}^{l} = E_{l-1} \\ |f_{ref}(i+vi, j+vj) - f_{cur}(i, j-1)| & \text{if} \quad E_{dir}^{l} = E_{l} \\ |f_{ref}(i+vi, j+vj) - f_{cur}(i+1, j-1)| & \text{if} \quad E_{dir}^{l} = E_{l+1} \end{bmatrix} \quad (10)$$

$$E_{l}^{left} = \begin{bmatrix} |f_{ref}(i+vi, j+vj) - f_{cur}(i-1, j-1)| & \text{if} \quad E_{dir}^{l} = E_{l-1} \\ |f_{ref}(i+vi, j+vj) - f_{cur}(i-1, j)| & \text{if} \quad E_{dir}^{l} = E_{l} \\ |f_{ref}(i+vi, j+vj) - f_{cur}(i-1, j+1)| & \text{if} \quad E_{dir}^{l} = E_{l+1} \end{bmatrix} \quad (11)$$

$$E_{l}^{bot} = \begin{bmatrix} |f_{ref}(i+vi, j+vj+S-1) - f_{cur}(i-1, j+S)| & \text{if} \quad E_{dir}^{l} = E_{l-1} \\ |f_{ref}(i+vi, j+vj+S-1) - f_{cur}(i, j+S)| & \text{if} \quad E_{dir}^{l} = E_{l} \\ |f_{ref}(i+vi, j+vj+S-1) - f_{cur}(i+1, j+S)| & \text{if} \quad E_{dir}^{l} = E_{l+1} \end{bmatrix} \quad (12)$$

$$E_{l}^{right} = \begin{bmatrix} |f_{ref}(i+vi+S-1, j+vj) - f_{cur}(i+S, j-1)| & \text{if} \quad E_{dir}^{l} = E_{l-1} \\ |f_{ref}(i+vi+S-1, j+vj) - f_{cur}(i+S, j)| & \text{if} \quad E_{dir}^{l} = E_{l} \\ |f_{ref}(i+vi+S-1, j+vj) - f_{cur}(i+S, j+1)| & \text{if} \quad E_{dir}^{l} = E_{l+1} \end{bmatrix} \quad (13)$$

$$DTBMC_{position} = w_{position} \times \sum_{n=0}^{S-1} E_{n}^{position} \qquad position \in \{top, left, bottom, right\} \quad (14)$$

where $E_{l}^{top}$, $E_{l}^{left}$, $E_{l}^{bot}$ and $E_{l}^{right}$ are absolute differences for each pixel in the determined direction in top, left, bottom, and right boundary, respectively. In Eq. (14), $DTBMC_{position}$ is used for calculating the direction distortion for each boundary.

### 2.3 Adaptive Boundary Matching Criterion (ABMC)

The proposed algorithm uses two explained boundary matching criterions, adaptively. In this algorithm, both OBMC and DTBMC are calculated for each boundary of the candidate MB. The block diagram of the proposed algorithm is shown in Fig. 3. According to this figure, after getting the coordinate of damaged MB, a set of candidate MVs is selected. This set includes adjacent MVs of damaged MB, mean and median MVs of them, zero MV, and the corresponding MV of damaged MV of the last frame.

It is assumed in ABMC that the criterion with the least boundary distortion can reconstruct the edge in the boundary of the damaged MB. So, for each boundary of the candidate MBs, one of the given boundary distortion criterions is selected. Finally, the candidate MV with the least boundary distortion is selected as the MV of damaged MB.

$$ABMC = \frac{\arg\min}{position}\langle OBMC_{position}, DTBMC_{position}\rangle \quad (15)$$

$position \in \{top, left, bottom, right\}$

### 2.4 Weight Determination

As already mentioned, $w$ is the weight of each outer boundary of the damaged MB. In this section, the weight determination for outer boundaries of the damaged MB is explained (see Fig. 4).

When the adjacent MB is completely correct, this weight is equal to one. When there is no adjacent MB or it is damaged, this weight is equal to zero. Moreover, if the adjacent MB is the result of error concealment from previous stages, according to the estimation accuracy of adjacent MB, this weight is between zero and one.

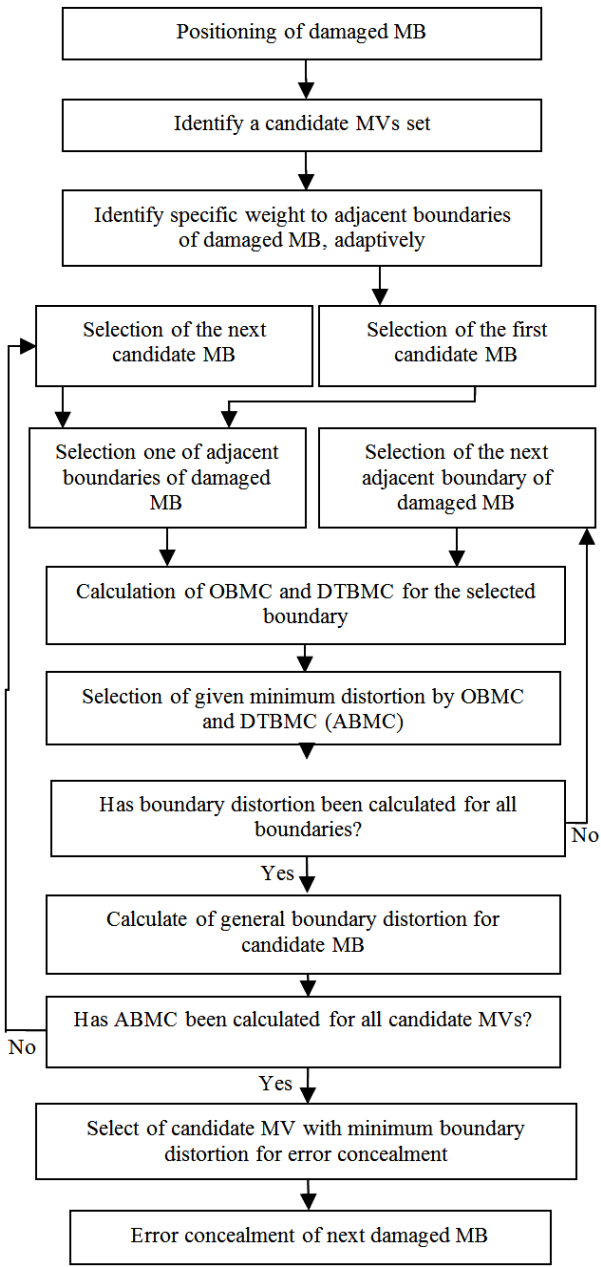

Fig. 3 Block diagram of the proposed algorithm.

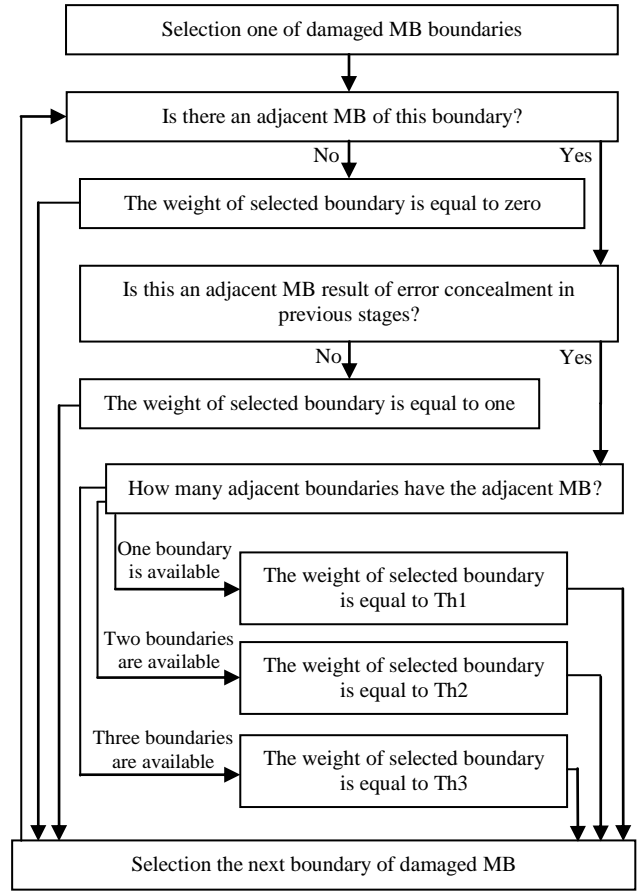

Fig. 4 Block diagram of weight determination for each outer boundaries of damaged MB.

If one of adjacent MBs is the result of previous error concealment stages, the proposed algorithm gives different weights to each adjacent boundary. In this case, at first, the numbers of surround MBs of adjacent MBs are determined. Then, according to the given numbers, one of Th1, Th2, or Th3 is determined for each outer boundary. Here, the values of Th1, Th2, and Th3 are 0.5, 0.7, and 0.9, respectively.

## 3  Experimental Results

In order to evaluate the performance of the proposed algorithm, various types of CIF (325×288) video test sequences including "Mother and Daughter", "Bus", "Stefan", "Foreman", "Paris", "Coastguard", "Vectra", and "Silent", and also QCIF (176×144) video test sequences including "Mother and Daughter", "Bus", "Foreman", "Car phone", "Miss America", "Football", "Bridge-close", and "Suzie" are used. These test sequences have different motions in consecutive frames.

The video test sequences are encoded in 4:2:0 format. The size of an MB is 16×16. To calculate MVs, the block matching algorithm [41] with exhaustive search (full search) and the search parameter equal to 7 ($p = 7$) was used. However, instead of using the full search method to estimate MVs, it can easily use any other known method, such as [42, 43]. The error in video frames with MB missing rate of 5 %, 10 %, and 20 % in each frame are created, randomly.

To evaluate the performance of the proposed algorithm, the damaged MVs are recovered by different algorithms. The performance of the proposed algorithm is compared with that of Average Motion Vector (AMV), BMA, DBM, DTBMA, OBMA, and DTEC methods. For increasing the accuracy of the results, the experiments for every algorithm are performed 20 times and the average values are used as the final result.

Fig. 5 shows the average PSNR values for reconstructed CIF test sequences of "Mother and Daughter", "Bus", "Stefan", "Foreman", "Paris", "Coastguard", "Vectra", and "Silent" with respect to the number frames, respectively. Also, Fig. 6 shows the average PSNR values for the reconstructed QCIF test sequences of "Mother and Daughter", "Bus", "Foreman", "Car phone", "Miss America", "Football", "Bridge-close", and "Suzie" with respect to the number frames, respectively. Experimental results obtained on these figures show that the proposed algorithm increases the average PSNR for test sequences in some frames about 5.20, 5.78, 5.88, 4.37, 4.41, and 3.50 dB compared to AMV, BMA, DBM, DTBMA, OBMA, and DTEC methods, respectively. Although other algorithms when run on different sequences may lose their efficiency, the experimental results show that the proposed algorithm maintains its efficiency in most examples.

The average PSNR values of luminance for 60 frames of CIF sequences and 70 frames of QCIF sequences are listed in Table 1 and Table 3, respectively.

To compare the quality and time complexity of various concealment algorithms the same PC (Intel Core i5, 2.4 GHz) is used. The average reconstruction time (in msec) for 60 frames of CIF sequences and 70 frames of QCIF sequences with average MB missing rate 10 % is listed in Table 2.

According to Tables 1 and 3, the proposed algorithm yields higher average PSNR performance than AMV, BMA, DBM, DTBMA, OBMA, and DTEC in both CIF and QCIF resolutions. From Table 1, one can observe that the proposed algorithm in CIF resolution is able to provide up to 2.1362, 1.5254, 1.5631, 1.4164, 1.2182, and 0.741 dB higher PSNR than the AMV, BMA, DBM, DTBMA, OBMA, and DTEC, respectively. Also, according to Table 3, the proposed algorithm in QCIF resolution is able to provide up to 1.9225, 1.9923, 1.9822, 1.5834, 0.922, and 0.3779 dB higher PSNR than the AMV, BMA, DBM, DTBMA, OBMA, and DTEC respectively.

Also, from Tables 1 to 3, one can observe that the proposed algorithm can estimate MVs of the damaged MBs without any considerable increase in computational complexity compared to AMV, BMA, DBM, DTBMA, OBMA, and DTEC. The proposed algorithm with adaptive selection from one of the weighted boundary matching criterions for each boundary of the candidate MBs, can achieve more accuracy in damaged MV estimation. This algorithm can work well, when faced with oblique edges and changing of edge direction in consecutive frames. So, it causes high objective and subjective quality in reconstructed frames.

Fig. 7(a) shows an error-free frame in CIF test sequence "Stefan" (Frame 8). Fig. 7(b) shows the damaged frame with average MB missing rate of 20 %. The reconstructed frames using the AMV, BMA, DBM, DTBMA, OBMA, DTEC, and the proposed algorithm are shown in Figs. 7(c), 7(d), 7(e), 7(f), 7(g), 7(h), and 7(i), respectively. Also, Fig. 8(a) shows an error-free frame in QCIF test sequence "Bus" (Frame 13). Fig. 8(b) shows the damaged frame with average MB missing rate of 20 %. Reconstructed frames using the AMV, BMA, DBM, DTBMA, OBMA, DTEC, and the proposed algorithm are shown in Figs. 8(c), 8(d), 8(e), 8(f), 8(g), 8(h), and 8(i), respectively. As observed from these figures, the proposed algorithm achieves higher subjective quality over other methods.

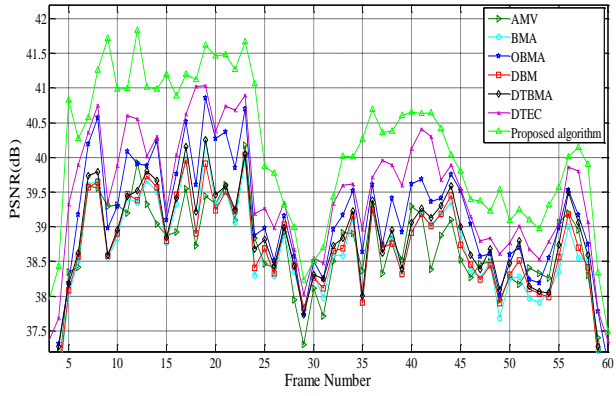
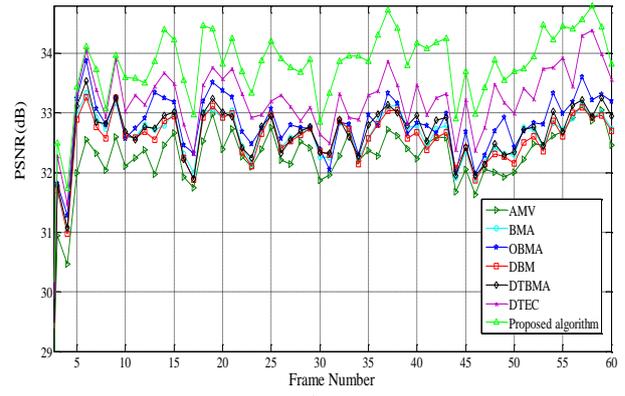
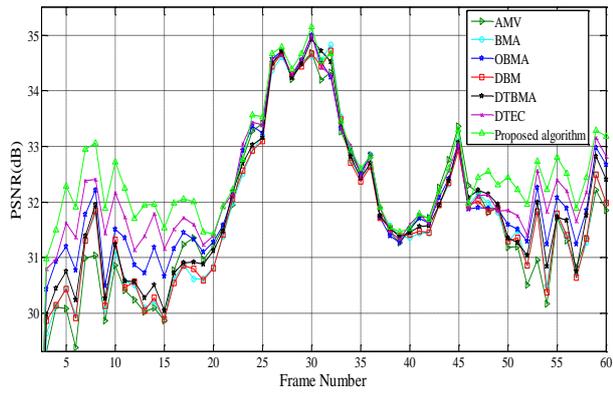
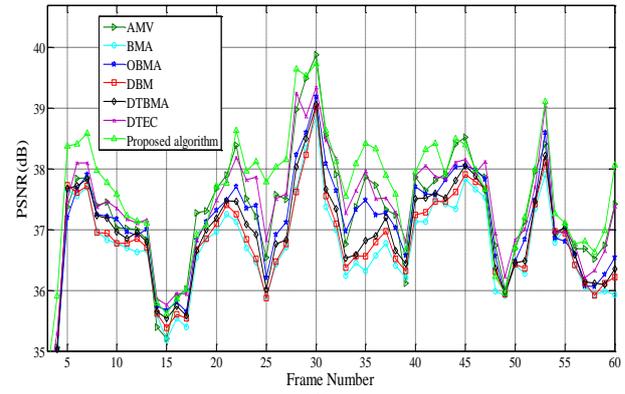
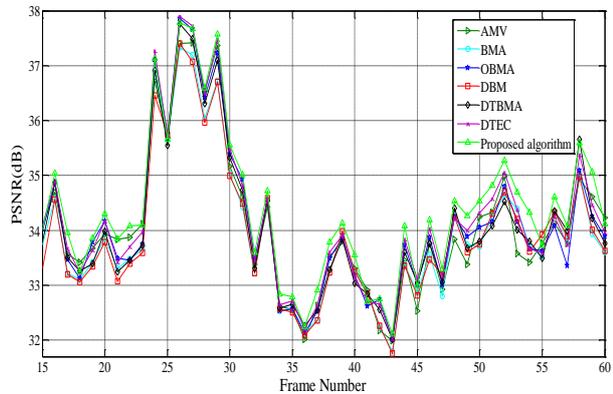
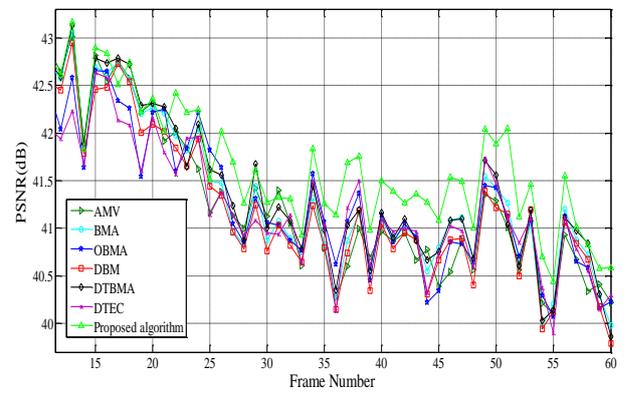
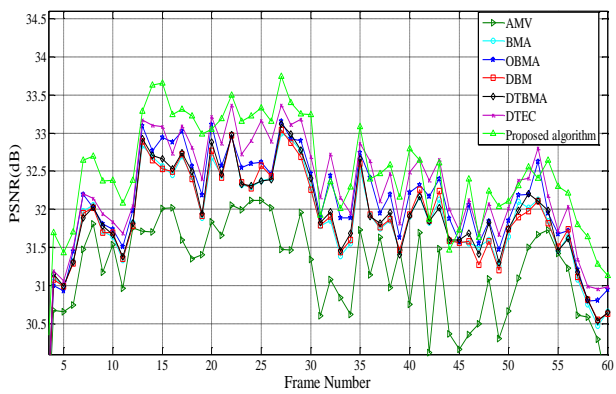
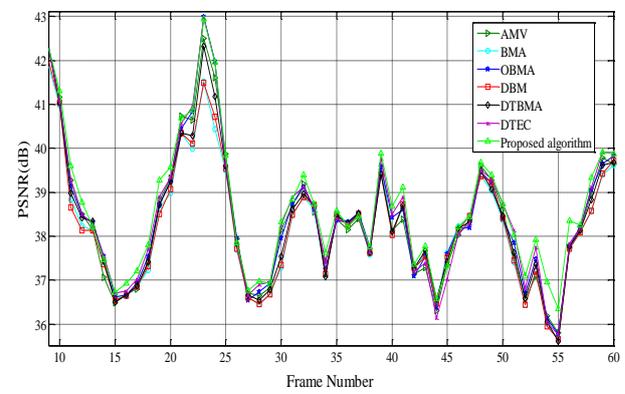

**Fig. 5** PSNR values of different CIF video test sequences with average MB missing rate of 20%: (a) "Mother and Daughter", (b) "Bus", (c) "Stefan", (d) "Foreman", (e) "Paris", (f) "Coastguard", (g) "Vectra", (h) "Silent".

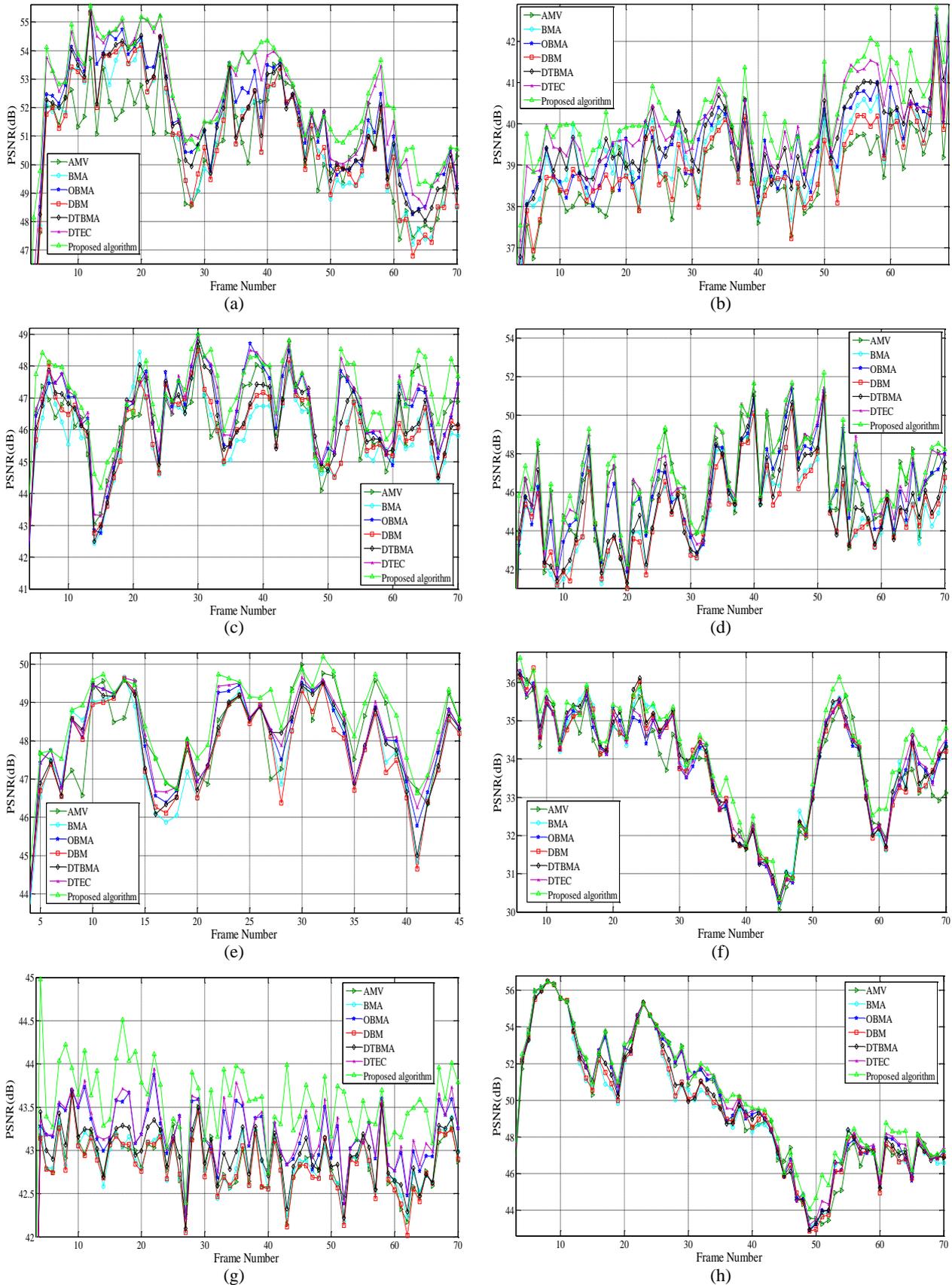

**Fig. 6** PSNR values of different QCIF video test sequences with average MB missing rate of 20%: (a) "Mother and Daughter", (b) "Bus", (c) "Foreman", (d) "Car phone", (e) "Miss America", (f) "Football", (g) "Bridge-close", (h) "Suzie".

**Table 1** Average PSNR (dB) for CIF video test sequences with different error concealment algorithms.

| CIF video sequence | Average MB missing rate | Error concealment algorithm | | | | | | |
|---|---|---|---|---|---|---|---|---|
| | | AMV | BMA | DBM | DTBMA | OBMA | DTEC | Proposed algorithm |
| **Mother and Daughter** | 5% | 42.9837 | 43.2544 | 43.3409 | 43.4565 | 43.8851 | 44.1487 | 44.3766 |
| | 10% | 40.6403 | 40.6146 | 40.6834 | 40.8011 | 41.1570 | 41.4326 | 41.9278 |
| | 20% | 38.0587 | 38.0603 | 38.0912 | 38.2165 | 38.5031 | 38.8730 | 39.4841 |
| **Bus** | 5% | 37.4828 | 38.3638 | 38.3149 | 38.6077 | 38.8291 | 39.2102 | 39.6190 |
| | 10% | 34.5265 | 35.0362 | 34.9985 | 35.1452 | 35.3434 | 35.8206 | 36.5616 |
| | 20% | 31.9168 | 32.2440 | 32.1967 | 32.2907 | 32.4389 | 32.8406 | 33.4174 |
| **Stefan** | 5% | 3.2865 | 37.4683 | 37.4119 | 37.6181 | 37.9375 | 38.0451 | 38.1507 |
| | 10% | 34.3860 | 34.4087 | 34.4209 | 34.5357 | 34.7296 | 34.9114 | 35.1206 |
| | 20% | 31.3703 | 31.3952 | 31.3953 | 31.5190 | 31.6977 | 31.9036 | 32.1549 |
| **Foreman** | 5% | 41.9488 | 41.5311 | 41.6532 | 41.8990 | 42.2671 | 42.5125 | 42.5952 |
| | 10% | 39.3925 | 38.7882 | 38.9033 | 39.0538 | 39.2125 | 39.4751 | 39.7477 |
| | 20% | 36.7484 | 36.1906 | 36.2875 | 36.4109 | 36.5518 | 36.8284 | 37.0602 |
| **Paris** | 5% | 40.4550 | 40.4277 | 40.3759 | 40.7215 | 40.9596 | 41.0815 | 41.2527 |
| | 10% | 37.5037 | 37.3697 | 37.3605 | 37.5258 | 37.6658 | 37.7511 | 37.9307 |
| | 20% | 33.6774 | 33.5154 | 33.4976 | 33.6524 | 33.7011 | 33.8098 | 33.9735 |
| **Coastguard** | 5% | 46.6251 | 46.3219 | 46.1923 | 46.4843 | 46.6727 | 46.7524 | 46.8285 |
| | 10% | 43.6241 | 43.6068 | 43.4842 | 43.6711 | 43.5701 | 43.5702 | 43.9792 |
| | 20% | 40.7782 | 40.8160 | 40.7048 | 40.8619 | 40.7207 | 40.6833 | 41.0999 |
| **Silent** | 5% | 43.9796 | 43.9866 | 44.0027 | 44.2457 | 44.4287 | 44.4970 | 44.5803 |
| | 10% | 41.3156 | 41.1055 | 41.1271 | 41.2589 | 41.3818 | 41.4304 | 41.5998 |
| | 20% | 38.4336 | 38.2926 | 38.3034 | 38.3963 | 38.4793 | 38.5353 | 38.6682 |
| **Vectra** | 5% | 35.9067 | 36.7310 | 36.7332 | 36.8271 | 37.1085 | 37.2719 | 37.4007 |
| | 10% | 33.4501 | 34.0174 | 34.0303 | 34.0714 | 34.2231 | 34.3680 | 34.7031 |
| | 20% | 30.8139 | 31.4630 | 31.4709 | 31.5046 | 31.6645 | 31.8487 | 32.0752 |

**Table 2** Average reconstruction time (msec) per MB for average MB missing rate 10%.

| Video resolution | Video sequence | Error concealment algorithm | | | | | | |
|---|---|---|---|---|---|---|---|---|
| | | AMV | BMA | DBM | DTBMA | OBMA | DTEC | Proposed algorithm |
| **CIF** | Mother and Daughter | 4.0284 | 4.1447 | 4.7149 | 4.7639 | 4.2026 | 4.1098 | 5.1688 |
| | Bus | 4.0225 | 4.1093 | 4.5900 | 4.6154 | 4.0762 | 4.1630 | 5.1209 |
| | Stefan | 4.0101 | 4.0379 | 4.6999 | 4.7422 | 4.0429 | 4.0864 | 5.1554 |
| | Foreman | 4.2793 | 4.3598 | 4.8814 | 4.9202 | 4.3474 | 4.3194 | 5.5031 |
| | Paris | 4.4469 | 4.4771 | 5.0368 | 4.9945 | 4.4623 | 4.5189 | 5.6717 |
| | Coastguard | 4.2809 | 4.3633 | 4.8780 | 4.9116 | 4.3606 | 4.2420 | 5.4997 |
| | Silent | 4.2860 | 4.4098 | 4.8960 | 5.0036 | 4.4071 | 4.4356 | 5.5325 |
| | Vectra | 4.1334 | 4.2275 | 4.7169 | 4.7691 | 4.2211 | 4.1935 | 5.3390 |
| **QCIF** | Mother and Daughter | 4.4132 | 4.5274 | 5.0062 | 5.0419 | 4.5257 | 4.5203 | 5.5926 |
| | Bus | 4.7392 | 4.8446 | 5.3585 | 5.4151 | 4.8427 | 4.7863 | 5.9960 |
| | Foreman | 4.5590 | 4.6766 | 5.1357 | 5.1911 | 4.6404 | 4.5654 | 5.7261 |
| | Car phone | 4.3683 | 4.4351 | 4.9785 | 5.0388 | 4.4297 | 4.9314 | 5.5630 |
| | Miss America | 4.2839 | 4.3319 | 4.9702 | 4.9806 | 4.3369 | 4.4077 | 5.4635 |
| | Football | 4.1736 | 4.3139 | 4.8869 | 4.9394 | 4.3062 | 4.2780 | 5.3532 |
| | Suzie | 4.5031 | 4.5801 | 5.1088 | 5.1570 | 4.5402 | 4.5651 | 5.7085 |
| | Bridge-close | 4.7785 | 4.8064 | 4.4140 | 5.4432 | 4.8043 | 4.8514 | 5.9341 |

**Table 3** Average PSNR (dB) for QCIF video test sequences with different error concealment algorithms.

| QCIF video sequence | Average MB missing rate | Error concealment algorithm | | | | | | |
|---|---|---|---|---|---|---|---|---|
| | | AMV | BMA | DBM | DTBMA | OBMA | DTEC | Proposed algorithm |
| Mother and Daughter | 5% | 60.3195 | 60.3675 | 60.4374 | 60.7488 | 60.9820 | 61.6572 | 61.8837 |
| | 10% | 54.5433 | 54.8156 | 54.8420 | 55.1186 | 55.2905 | 55.7455 | 55.9686 |
| | 20% | 50.0727 | 50.3747 | 50.3875 | 50.7894 | 50.9769 | 51.5161 | 51.8387 |
| Bus | 5% | 45.9344 | 46.6517 | 46.6089 | 46.8229 | 47.1047 | 47.2984 | 47.5509 |
| | 10% | 43.2557 | 44.2095 | 44.0861 | 44.3975 | 44.6449 | 44.8665 | 45.1782 |
| | 20% | 38.4128 | 38.8787 | 38.5669 | 39.2189 | 39.0350 | 39.5095 | 39.8597 |
| Foreman | 5% | 57.3576 | 57.2427 | 57.5092 | 57.8273 | 58.1297 | 58.1639 | 58.1755 |
| | 10% | 50.1606 | 49.7582 | 50.0023 | 50.3042 | 50.6794 | 50.8178 | 50.9351 |
| | 20% | 45.5803 | 45.1969 | 45.3654 | 45.5906 | 45.8327 | 45.9853 | 46.3632 |
| Car phone | 5% | 55.3566 | 54.9117 | 54.9113 | 55.3789 | 56.2635 | 56.1620 | 56.3773 |
| | 10% | 50.3072 | 49.7295 | 49.7803 | 50.2691 | 50.8754 | 51.2681 | 51.4319 |
| | 20% | 45.6800 | 44.4544 | 44.4645 | 44.8633 | 45.5247 | 46.1635 | 46.4467 |
| Miss America | 5% | 56.8242 | 56.8479 | 56.7860 | 56.9572 | 56.9871 | 57.0156 | 57.2858 |
| | 10% | 51.3385 | 51.0459 | 51.0382 | 51.2617 | 51.3808 | 51.4947 | 51.6408 |
| | 20% | 46.8870 | 46.6148 | 46.6341 | 46.8187 | 46.8981 | 47.0550 | 47.4058 |
| Football | 5% | 42.1522 | 42.7034 | 42.6054 | 42.8046 | 42.7745 | 42.8862 | 43.0610 |
| | 10% | 38.0890 | 38.3840 | 38.3379 | 38.5178 | 38.5026 | 38.5403 | 38.8155 |
| | 20% | 33.4608 | 33.6262 | 33.5984 | 33.7016 | 33.5662 | 33.6434 | 33.9442 |
| Suzie | 5% | 58.0402 | 58.2173 | 58.2584 | 58.3140 | 58.5761 | 58.7016 | 58.8653 |
| | 10% | 53.1561 | 52.7702 | 52.8307 | 52.9293 | 53.1680 | 53.3963 | 53.6067 |
| | 20% | 49.1793 | 48.8893 | 49.2544 | 48.8414 | 48.0532 | 49.4249 | 49.6070 |
| Bridge-close | 5% | 49.5497 | 49.4010 | 49.3613 | 49.4506 | 49.4834 | 49.5093 | 49.8213 |
| | 10% | 46.2310 | 46.2607 | 46.2416 | 46.2706 | 46.4934 | 46.5078 | 46.7539 |
| | 20% | 42.3552 | 42.3594 | 42.3136 | 42.4521 | 42.6488 | 42.6996 | 43.0565 |

## 4 Conclusion

In this paper, an adaptive boundary matching algorithm for temporal error concealment of video frames is presented. It uses two boundary matching criterions in order to increase the accuracy in damaged MV estimation. In the proposed algorithm, when the edges in boundaries of damaged MB are estimated correctly by DTBMC, this criterion is used; otherwise the OBMC is selected. It causes the proposed algorithm to overcome the difficulties of OBMA and DTBMA methods. Furthermore, considering the accuracy coefficient for each outer boundary of damaged MB can cause more accuracy and low error propagation. Experimental results show that the proposed algorithm increases the average PSNR for video test sequences in some frames about 5.20, 5.78, 5.88, 4.37, 4.41, and 3.50 dB compared to AMV, BMA, DBM, DTBMA, OBMA, and DTEC, respectively. Also, without any considerable computational complexity, it improves the objective and subjective qualities.

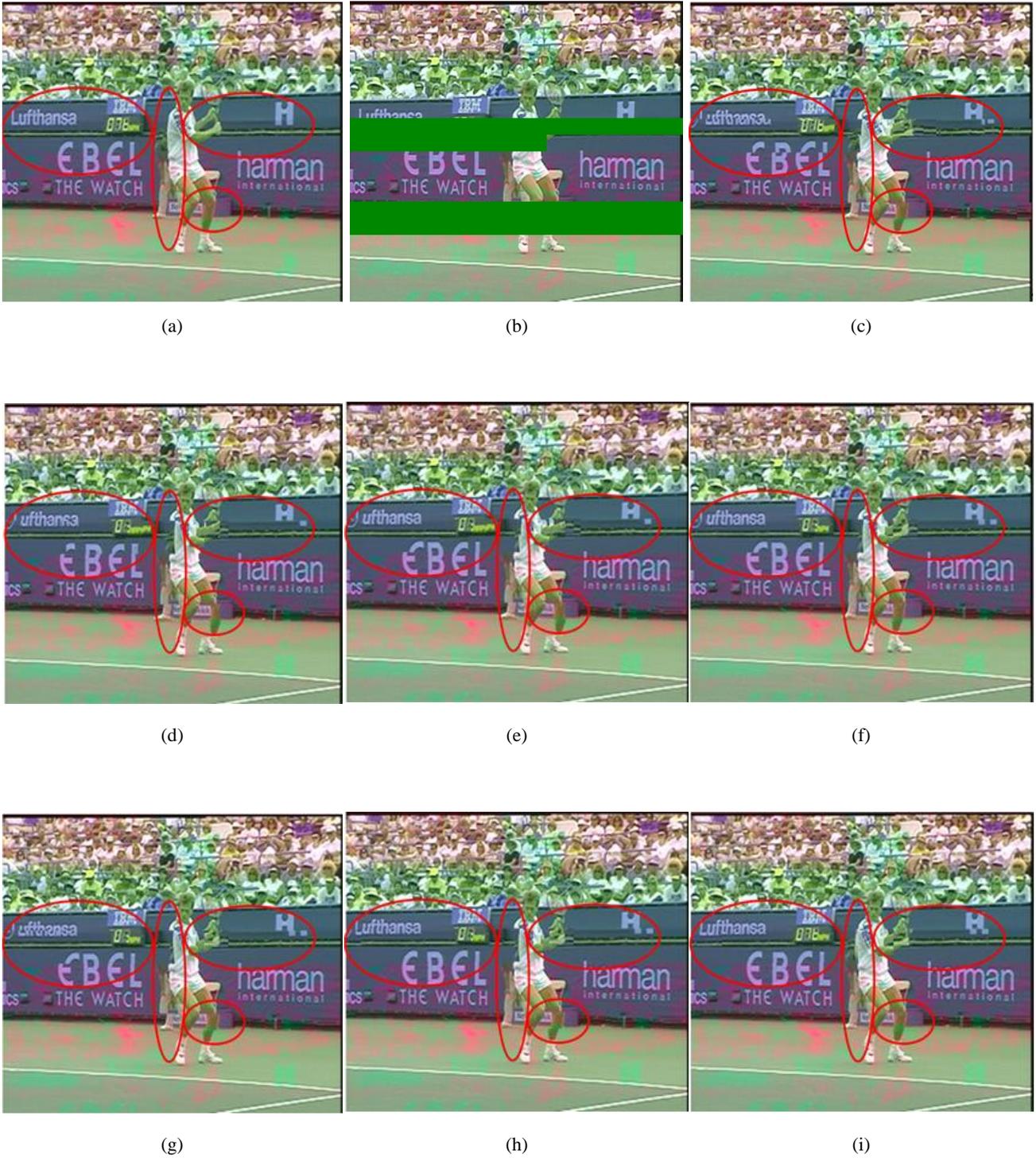

**Fig. 7** Subjective video quality comparison for Frame 8 in CIF video test sequence "Stefan" and average MB missing rate of 20%. (a): Free-error frame; (b): Corrupted frame; (c): Reconstructed frame using AMV; (d): Reconstructed frame using BMA; (e): Reconstructed frame using DBM; (f): Reconstructed frame using DTBMA; (g): Reconstructed frame using OBMA; (h): Reconstructed frame using DTEC; (i): Reconstructed frame using the proposed algorithm.

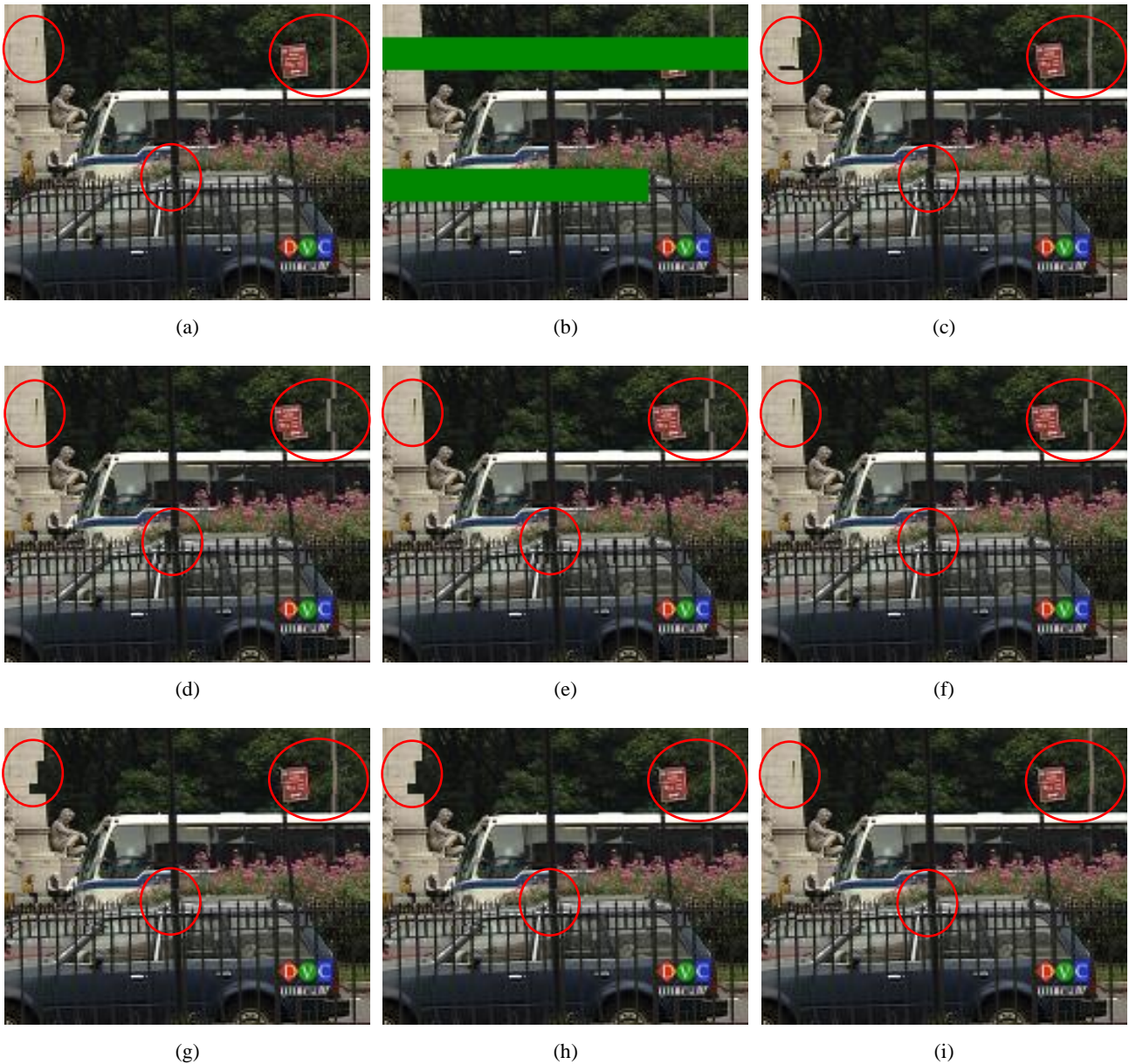

**Fig. 8** Subjective video quality comparison for Frame 13 in QCIF video test sequence "Bus" and average MB missing rate of 20%. (a): Free-error frame; (b): Corrupted frame; (c): Reconstructed frame using AMV; (d): Reconstructed frame using BMA; (e): Reconstructed frame using DBM; (f): Reconstructed frame using DTBMA; (g): Reconstructed frame using OBMA; (h): Reconstructed frame using DTEC; (i): Reconstructed frame using the proposed algorithm.

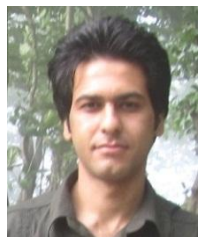

**Seyed Mojtaba Marvasti-Zadeh** was born in Yazd, Iran, on July 15th 1986. He received the Associate's degree in electronics from Technical Faculty of Imam Ali, Yazd, Iran, in 2008. Then, he received the B.Sc. degree in electrical engineering from the Department of Electrical and Electronic Engineering, Science and Arts University, Iran, in 2010, and the M.Sc. degree from the Department of Electrical and Computer Engineering, Yazd University, Iran, in 2014. He was awarded as the best graduate student of Electrical and Electronic Departments, Technical Faculty of Imam Ali and Science and Arts University, in 2008 and 2010, respectively. His research interests include image and video signal processing, video error concealment, computer vision and multimedia communications.

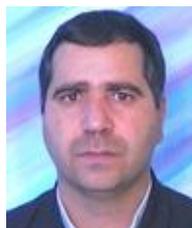

**Hossein Ghanei-Yakhdan** was born in Yazd, Iran in 1966. He received the B.Sc. degree in electrical engineering from Isfahan University of Technology in 1989, the M.Sc. degree in 1993 from K. N. Toosi University of Technology and the Ph.D degree in 2009 from Ferdowsi University of Mashhad. Since 1994, he has been with the Department of Electrical and Computer Engineering, Yazd University. His research interests are in digital video and image processing, error concealment and error-resilient coding for video communication and digital image watermarking.

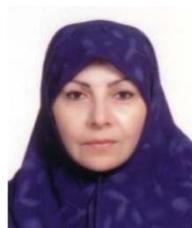

**Shohreh Kasaei** (M'05-SM'07) received her B.Sc. degree from the Department of Electronics, Faculty of Electrical and Computer Engineering, Isfahan University of Technology, Iran, in 1986, her M.Sc. degree from the Graduate School of Engineering, Department of Electrical and Electronic Engineering, University of the Ryukyus, Japan, in 1994, and the Ph.D. degree from Signal Processing Research Centre, School of Electrical and Electronic Systems Engineering, Queensland University of Technology, Australia, in 1998. She joined Sharif University of Technology since 1999, where she is currently a full professor and the director of Image Processing Laboratory (IPL). Her research interests include 3D computer vision, 3D object tracking, human activity recognition, multi-resolution texture analysis, scalable video coding, image retrieval, video indexing, face recognition, hyperspectral change detection, video restoration, and fingerprint authentication.